\documentclass[12pt]{article}
\usepackage{amssymb,slashed}
\usepackage{amsmath,euscript}
\usepackage{hyperref}
\usepackage{cancel}

\topmargin
-1.5cm
\textwidth
15.5cm
\textheight
23.5cm
\oddsidemargin
0.7cm
\evensidemargin
1.2cm
\begin{document}
\begin{center}
\LARGE{\bf New F-theory lifts II\,:\\ \vskip1.5mm Permutation orientifolds  and\\ enhanced singularities  \\[15mm]}
\large{Andr\'es Collinucci\\[3mm]}
\footnotesize{Insitute for Theoretical Physics, Vienna University of Technology\\
Wiedner Hauptstr. 8-10, 1040 Vienna, Austria}\\
\vskip 1.5cm
\small{\bf Abstract} \\[5mm]
\end{center}
\begin{center}
\begin{minipage}[h]{16.0cm}

In this paper, a procedure is developed to construct compact F-theory fourfolds corresponding to perturbative IIB O7/O3 models on CICY threefolds with permutation involutions. The method is explained in generality, and then applied to specific examples where the involution permutes two Del Pezzo surfaces. The fourfold construction is successfully tested by comparing the D3 charges predicted by F-theory and IIB string theory.

The constructed fourfolds are then taken to the locus in moduli space where they have enhanced SU(5) singularities. A general, intuitive method is developed for engineering the desired singularities in Weierstrass models for complicated D7-brane setups.

\end{minipage}
\end{center}
\newpage

\tableofcontents
\newpage
\section{Introduction}
Over the past year, the field has seen progress in F-theory model building delivered at a very high pace. This renewed interest is motivated in part by the fact that F-theory enjoys some of the advantages of heterotic GUT model building, such as the presence of exceptional gauge groups, while it avoids some of its pitfalls, such as necessarily making the hypercharge U(1) massive when the GUT group is broken by fluxes. 
The other feature that makes F-theory attractive, is that it lends itself to `local' model building. These points are summarized in \cite{Wijnholt:2008db}, and explained in the key papers that have triggered this progress, \cite{Beasley:2008dc, Beasley:2008kw, Donagi:2008ca, Donagi:2008kj}. The long list of recent papers on local F-theory models includes \cite{Tatar:2008zj, Tatar:2009jk, Font:2008id, Jiang:2009za, Li:2009cy, Jiang:2008yf, Chen:2009me}.

However, one ultimately needs global information about the compact CY fourfold. Some of the pitfalls of local model building are explained in \cite{Andreas:2009uf}. For instance one cannot engineer the intersections of 7-branes at will, as the global structure of the CY fourfold may impose or forbid gauge group enhancement loci on the GUT brane both at codimension two and three in the ambient, base threefold. In \cite{Witten:pitp}, it was also mentioned that a desired IIB setup might simply not admit a corresponding axio-dilaton solution.

Since there has only recently been a shift in the literature towards global F-theory model building (see for instance \cite{Marsano:2009ym, Donagi:2009ra}), not many examples of non-CY threefolds have been studied in the F-theory literature that contain the phenomenologically desirable shrinkable surfaces. On the other hand, numerous examples of CY threefolds in with shrinkable surfaces have been constructed for the sake of making perturbative IIB orientifold models. In \cite{Collinucci:2008sq}, CY threefolds were constructed with three exceptional divisors. Such divisors can be blown down to curves or points. In \cite{Blumenhagen:2008zz}, CY threefolds were constructed with surfaces that satisfy the even more stringent condition of being Del Pezzo.

Although the progress in F-theory model building enables to some extent a departure from perturbative string theory, IIB perturbative models have their advantages. For instance, one can compute $\alpha'$ corrections, which are crucial for closed string moduli stabilization in the LARGE volume scenario \cite{Balasubramanian:2005zx, Conlon:2005jm, Cicoli:2008va}. Another advantage is the fact that DBI fluxes on D7-branes can be explicitly constructed much more easily than can four-form fluxes in global F-theory models.

Hence, it would prove very useful to be able to construct F-theory lifts for all IIB orientifold models. This would provide such models with the consistency/existence checks mentioned above, plus additional information, such as the possibility to generate non-perturbative superpotentials via Euclidean D3-branes \cite{Witten:1996bn}.
\vskip 2mm
In \cite{Collinucci:2008zs}, a procedure was developed to construct F-theory lifts for IIB orientifold models with holomorphic involutions of the form $x \rightarrow -x$. This entails orbifolding the CY threefold of the IIB setup with respect to the involution. Although many F-theory models corresponding to such orientifold setups existed before, the only cases under control involved elliptically fibered CY fourfolds of which the threefold base of is toric. A typical example being the octic hypersurface CY threefold in $\mathbb{P}^4_{1,1,1,1,4}$, of which the $\mathbb{Z}_2$ orbifold is simply $\mathbb{P}^3$. Requiring the base threefold to be toric is a severe limitation for model building. For instance, this limitation means one cannot lift a single IIB orientifold model on the CY quintic!
The key result in \cite{Collinucci:2008zs} shows how to construct base threefolds as complete intersections in toric spaces, starting from a CY threefold and an involution of the aforementioned type.

Involutions of the form $x \rightarrow -x$ necessarily have $h^{1,1}_-=0$, i.e. act trivially on the homology of the CY threefold. In this paper, a technique is developed to construct orbifolds of CY threefolds with respect to involutions that permute cycles, i.e. with $h^{1,1}_- \neq 0$. This will open up the possibility to create F-theory lifts for more general and much more interesting scenarios.
\vskip 2mm
The purpose of this paper is twofold: A general technique for orbifolding CY threefolds will be exposed, in order to be able to create F-theory lifts for IIB O7/O3 permutation orientifold models with a single generic D7-brane. 

The second purpose of the paper is to show how to engineer the singularity enhancements in order to create GUT-like models, such as SU(5) models, by using the rules laid out in \cite{Collinucci:2008pf} for orientifold invariant D7-branes. Although one can in principle engineer such singularities by guessing the so-called \emph{Tate form} of the elliptic fibration, I will show how one can build a complete setup with several stacks in a constructive, intuitive way by exploiting Sen's limit of the \emph{Weierstrass form}. Once the Weierstrass form is obtained, the Tate form is easily deduced.

\vskip 2mm
This paper is organized as follows: In section \ref{sec:orbifolding}, the method for constructing orbifolds of CY threefolds with respect to permutation involutions is shown. This is first shown for the simple, two-dimensional example of $\big(\mathbb{P}^1 \times \mathbb{P}^1\big)/\mathbb{Z}_2\,$; then a general procedure is outlined; and finally, a specific CY threefold is treated where the involution exchanges two Del Pezzo surfaces.

In section \ref{sec:quasismooth}, I show how to construct the elliptically fibered F-theory fourfold for the specific example of the permutation orientifold of the previous section. The construction is explained in generality, and then for the example. This yields CY fourfolds at their generic locus in complex structure moduli space. Finally, the validity of the construction is submitted to a highly non-trivial check: The formula of \cite{Sethi:1996es} for the D3 tadpole in terms of the Euler characteristic of the fourfold is used. By using the K-theoretic techniques refined in \cite{Collinucci:2008pf}, the curvature-induced D3 tadpole for the IIB D7/O7 setup is compared to the Sethi-Vafa-Witten formula. The match is perfect.

In section \ref{sec:quasismooth}, we depart from the generic locus of the fourfold, and move on to the case with enhanced singularities. Although singularities can be engineered by using the Tate form of the elliptic fibration, a procedure is developed to construct stacks of D7-branes in a much more intuitive way by using the Weierstrass form, and the rule that D7-branes intersect O7-planes only at double-points, as elucidated in \cite{Collinucci:2008pf}. First, a lift is constructed for a simple SU(5) model with two `flavor' branes. Then, a more general SU(5) model is lifted. All relevant enhanced singularities are successfully detected.

In the appendix \ref{app}, a specific, three-generation model from \cite{Blumenhagen:2008zz} is lifted. 

\section{Orbifolding permutation involutions} \label{sec:orbifolding}
The first step to constructing an F-theory lift for a perturbative IIB model is constructing the base $B_3$ of the elliptically fibered CY fourfold $Y_4$. This base is obtained by orbifolding the CY threefold $X_3$ by the orientifold involution. In \cite{Collinucci:2008zs}, a technique was shown to create orbifolds of threefolds for holomorphic involutions with $h^{1,1}_-=0$, i.e. involutions that act trivially on the homology of $X_3$. Such involutions usually have the form $x \rightarrow -x$, for some coordinate $x$. 

In this section, I will develop a technique to orbifold spaces with respect to involutions with $h^{1,1}_- \neq 0$, i.e. involutions that permute homology cycles of $X_3$. First, I will treat the simple, two-dimensional case of $\big(\mathbb{P}^1 \times \mathbb{P}^1\big)/ \mathbb{Z}_2$, where the involution exchanges the two factors. This example will turn out to contain all the structure needed to extend the procedure to any toric space with any permutation involution that has a codimension one fixed-point locus (and possible higher codimension fixed-point loci).

Afterwards, I will summarize the general procedure, and then move on to a specific CY threefold with two Del Pezzo surfaces that are exchanged by the involution.

\subsection{$\big(\mathbb{P}^1 \times \mathbb{P}^1\big)/\mathbb{Z}_2$: The prototype orbifold} \label{subsec:prototype}
To gain insight into the general structure of holomorphic involutions that generate codimension one fixed-point loci and permute divisors, we now turn to the simplest possible complex, compact space that admits such an involution, namely the product space $X \equiv \mathbb{P}^1 \times \mathbb{P}^1$, with projective coordinates
\begin{align}
&(x_1, \, x_2;\, x_3,\, x_4) \sim (\lambda_1\, x_1,\, \lambda_1\, x_2;\, \lambda_2\, x_3,\, \lambda_2\, x_4)\,, \quad \lambda_1, \lambda_2 \in \mathbb{C}^*\,.
\end{align}
The second homology of this space is simply $H_2(X, \mathbb{Z}) = \mathbb{Z}^2$, generated by two divisor classes $D_A, D_B$, both of which admit $\mathbb{P}^1$'s as representatives.

The simplest conceivable permutation involution that acts non-trivially on the second homology of $X$ can be defined as the following map
\begin{equation} \label{invp1}
\sigma: \quad (x_1,\, x_2,\, x_3,\, x_4) \leftrightarrow  (x_3,\, x_4,\, x_1,\, x_2)\,.
\end{equation}
This map is a well-defined bijection. This involution simply exchanges the two divisor classes
\begin{equation}
\sigma_*: D_A  \leftrightarrow D_B\,.
\end{equation}
The fixed-point locus is the diagonally embedded $\mathbb{P}^1$, which is in the divisor class $D_A+D_B$, given by the set of points that satisfy
\begin{align}
(x_1, x_2) &= (\lambda\,x_3, \lambda\,x_4)\,, \quad {\rm for \ some} \quad \lambda \in \mathbb{C}^*\,.
\end{align}
This divisor can be described algebraically as the vanishing locus of the following equation
\begin{equation} \label{fpp1}
x_2\,x_3 - x_1\,x_4=0\,.
\end{equation}
Note, that this equation has the right bi-degree, i.e. $(1,1)$ under the two projective scalings, to give us a divisor of class $D_A+D_B$.
\vskip 2mm
The goal of this section is to orbifold $X$ by this involution. In order to do this, let us make a list of all possible sections of line bundles (i.e. homogeneous polynomials) that are eigenvectors under the involution \eqref{invp1}. These are all generated by the following sections:
\begin{equation} \label{sectionsp1}
(s_1, s_2, s_3, s_4) \equiv (x_1\,x_3\,, x_2\,x_4\,, x_1\,x_4+x_2\,x_3\,, x_1\,x_4-x_2\,x_3)\,,
\end{equation}
The involution acts on these generators as follows:
\begin{equation} \label{invonsectionsp1}
\sigma(s_i) = + s_i\,, \quad {\rm for} \quad i \neq 4\,, \quad {\rm and} \quad \sigma(s_4)=-s_4\,.
\end{equation}
To define the orbifold, we simply take the positive eigenvectors and define the following map into $\mathbb{P}^2$:
\begin{alignat}{2}
\left(\mathbb{P}^1\right)^2: (x_1, \ldots, x_4)\quad &\rightarrow \quad &\mathbb{P}^2: (y_1, y_2, y_3)&\nonumber \\
(x_1,\,x_2,\, x_3,\, x_4)\quad &\mapsto \quad &(y_1,\, y_2,\, y_3)& = (s_1, s_2, s_3)\,.
\end{alignat}
It can easily be checked that this map is surjective, and $2:1$ except at the fixed-point locus \eqref{fpp1}. Therefore, we conclude that
\begin{equation}
\fbox{$\big(\mathbb{P}^1 \times \mathbb{P}^1\big) / \mathbb{Z}_2 \cong \mathbb{P}^2$}
\end{equation}
\vskip 4mm
There is also a more indirect, but perhaps more intuitive approach to reach this conclusion. I will briefly explain it here. One can first use the sections $(s_1, \ldots, s_4)$ to define an embedding of $\mathbb{P}^1 \times \mathbb{P}^1$ into $\mathbb{P}^3$, as follows:
\begin{alignat}{2}
\left(\mathbb{P}^1\right)^2: (x_1, \ldots, x_4)\quad &\rightarrow \quad &\mathbb{P}^3: (y_1, y_2, y_3, y_4)&\nonumber \\
(x_1,\,x_2,\, x_3,\, x_4)\quad &\mapsto \quad &(y_1,\, y_2,\, y_3, y_4)& = (s_1, s_2, s_3, s_4)\,.
\end{alignat}
The image of the embedding is then defined algebraically as the hypersurface
\begin{equation} \label{quadricp1}
4\,y_1\,y_2 - y_3^2+y_4^2=0\,.
\end{equation}
This is known as a \emph{Segre embedding}. 
In this language, the involution acts on the ambient $\mathbb{P}^3$ via as $y_4 \rightarrow -y_4$.
In order to orbifold $X$, we first define an orbifold of the ambient space $\mathbb{P}^3$ as a weighted projective space as follows:
\begin{align}
o: \quad \mathbb{P}^3  &\longrightarrow \,\mathbb{P}^3_{1,1,1,2}\\
(y_1,\, y_2,\, y_3,\, y_4) &\longmapsto \,(z_1,\, z_2,\, z_3,\, z_4) =(y_1, y_2, y_3, y_4^2)\,,
\end{align}
where the $z_i$'s are homogeneous coordinates of the weighted projective space.
This map is 2 to 1 except at the fixed-point locus $y_4=0$, and at the point $(0,0,0,1)$, where it is one to one. The latter point, however, is not on the surface $X$ defined by \eqref{quadricp1}. We can now finally define the sought-for orbifold $X/\mathbb{Z}_2$ simply as a quadric hypersurface in the weighted projective space.
\begin{align}
&z_1\,z_2-z_3^2+z_4=0 \,, \quad \in \, \mathbb{P}^3_{1,1,1,2}\,.
\end{align}
Finally, we notice that we can eliminate the coordinate $z_4$ through this equation, since it appears linearly. This leaves us with $\mathbb{P}^2$, as expected.
\vskip 2mm
\subsection{General procedure}
The procedure illustrated in \ref{subsec:prototype} generalizes to more complicated situations rather straightforwardly. 

Let $X_3$ be a complete intersection CY threefold in a toric ambient space $T_d$ of dimension $d$, with a holomorphic involution $\sigma$ acting on $T_d$ that leaves $X_3$ invariant. Such an involution will act on the homogeneous coordinates either by inflicting minus signs on them, or by permuting sets of coordinates
\begin{align}
\sigma: x_i & \mapsto -x_i\,, \quad {\rm or}\\
\sigma: (x_1, \ldots, x_k) & \mapsto (x_{\pi(1)}, \ldots, x_{\pi(k)})\,.
\end{align}
In order to construct $X_3/\mathbb{Z}_2$, one first constructs $T_d/\mathbb{Z}_2$.
This is achieved as follows. Suppose that $\sigma$ acts non-trivially on the first $n$ coordinates of $T_d$. One then writes down the basis of sections of line bundles, $\{s_1, \ldots, s_N\}$, that are invariant under $\sigma$. Note, that $N$ could be greater than $n$, in which case the $s_i$ satisfy $N-n$ relations $r_i$. Then, one defines a new toric manifold, $T_{d-n+N}$, by replacing the $n$ coordinates with the $N$ sections, and leaving the other coordinates as they are. 
\begin{equation}
T_{d-n+N}\,:(s_1, \, \ldots,\, s_N,\, x_{n+1},\,\ldots)\,.
\end{equation}
The orbifold space $T_d/\mathbb{Z}_2$ is then defined as the complete intersection of the relations $r_i$ in the ambient space $T_{d-n+N}$
\begin{equation}
\bigcap_{i=1,\, \ldots,\, N-n} \, \{r_i=0\} \, \subset T_{d-n+N}\,.
\end{equation}
Now, one simply rewrites all equations that defined the original threefold $X_3$ in terms of these new, involution-invariant coordinates, and that will yield $X_3/ \mathbb{Z}_2$.
\vskip 2mm
In the case where the involution by permutes two pairs of coordinates as follows:
\begin{equation}
\sigma: (x_1,\, x_{2}\,, x_3,\, x_{4}, \ldots) \quad \mapsto \quad (x_3,\, x_{4}\,, x_1,\, x_{2}, \ldots)\,.
\end{equation}
One can directly define the orbifold of the ambient space $T_d$, one writes a `truncated Segre-like' map into a new toric space $\tilde T_d$  with homogeneous coordinates:
\begin{equation}
\tilde T_d: (y_1, y_2, y_3 \ldots)\,,
\end{equation}
by mapping the four coordinates that participate in the involution as follows: 
\begin{align}
(x_1,\, x_{2}\,, x_3,\, x_{4}) &\mapsto (y_1, y_2, y_3)\\
&=(\,x_1\, x_{3},\, x_{2}\, x_{4},\, x_{1}\,x_{4}+x_{2}\,x_{3})\,, \nonumber
\end{align}
and all other coordinates are mapped into equal coordinates.
Note, that this map will be consistent with all projective scalings by construction. 

In the even simpler case which was treated in \cite{Collinucci:2008zs}, where the involution simply gives a minus sign to one coordinate, $x_1 \rightarrow -x_1$, this amounts to defining a map into a new toric space as follows:
\begin{equation}
(x_1, x_2, \, \ldots) \mapsto (y_1, y_2,\, \ldots) =  (x_1^2, \, x_2,\, \ldots)
\end{equation}

\subsection{Preparing the base for a CY fourfold: \\ Orbifold of $Q^{(dP_7)^2}$} \label{subsec:baseqdp72}
In this section, I will apply the method described in the previous sections and in \cite{Collinucci:2008zs} to prepare the threefold base for an F-theory CY fourfold, based on a simple orientifold model on a CY threefold that admits a permutation involution. The purpose of this section is to illustrate and check the construction of the orbifold in a non-trivial example. The F-theory fourfold will subsequently be constructed in the next section.\\

The CY threefold in question was constructed in \cite{Blumenhagen:2008zz}, and is referred to as $Q^{(dP_7)^2}$. It is obtained by subjecting the quintic CY to two Del Pezzo transitions with the methods of \cite{Malyshev:2007yb} and \cite{Grimm:2008ed}. It can be regarded as a hypersurface, where the corresponding ambient space is $\mathbb{P}^4$ blown-up torically and sequentially at two points outside the threefold. The projective weights of the coordinates of the toric ambient space are given in table \ref{tab:qdp72}.

\begin{table}[h] 
\begin{centering} 
\begin{tabular}{|c|c|c|c|c|c|c||c|}  
\hline 
$x_{1}$ & $x_{2}$ & $x_{3}$ & $x_{4}$ & $x_{5}$ & $x_{6}$ & $x_7$ &$d$\tabularnewline
\hline
\hline 
$1$ &$1$ &$1$ &$1$ &$1$ &$0$ & $0$ & $5$\tabularnewline
\hline 
$0$ &$0$ &$0$ &$0$ &$1$ &$1$ &$0$ &2 \tabularnewline
\hline 
$0$ & $0$ &$0$ &$1$ &$0$& $0$ &$1$ &2 \tabularnewline
\hline
\end{tabular}
\par
\end{centering}
\caption{The rows indicate the projective weights of the coordinates under the two toric $\mathbb{C}^*$ actions for the $Q^{(dP_7)^2}$ space. The last column indicates the degree of a CY hypersurface.}
\label{tab:qdp72}
\end{table} 

The CY threefold is then a hypersurface given by a polynomial of multi-degree $(5,2,2)$.

This space admits several triangulations that allow for a smooth CY hypersurface. I will choose the one whose corresponding Stanley-Reisner ideal is\footnote{Here, each of the three entries of this list is a set of coordinates that are not allowed to vanish simultaneously.}
\begin{equation}
(x_1 x_2 x_3;\, x_5 x_6; x_4 x_7)\,.
\end{equation}
Defining $D_i$ as the divisor $x_i=0$, and choosing the basis $(D_1, D_6, D_7)$, the intersection numbers for this CY are\footnote{The coefficient in front of each term corresponds to the intersection number of the three divisors denoted by the term.}
\begin{align} \label{intqdp72}
I = &2\,(D_7^3+D_6^3)+2\,D_1^2\,(D_7+D_6)-D_7^2\,(2\,D_1+D_6) \nonumber\\
&-D_6^2\,(2\,D_1+D_7)+D_1\,D_6\,D_7\,.
\end{align}
By construction, this CY has two Del Pezzo surfaces. Namely, $D_6$ and $D_7$ are dP$_7$'s. 
\vskip 2mm
We may now define the following permutation involution
\begin{equation}
x_4 \leftrightarrow x_5\,, \quad x_6 \leftrightarrow x_7\,,
\end{equation}
which permutes the two dP surfaces.
Following the procedure described in the previous chapter, we can directly define the orbifold $T_4/\mathbb{Z}_2$ of the toric ambient $T_{4}$, by the following 2 to 1 map
\begin{align}
(x_4,\, x_5, \, x_6, \, x_7) \mapsto &(z_4, \, z_5, \,z_6) \\
= &\left(2\,x_4\,x_5, \, 2\,x_6\,x_7, \, x_5\,x_7+x_4\,x_6 \right)\,.
\end{align}
The toric data for $T_4/\mathbb{Z}_2$ is found in table \eqref{tab:qdp7/z2}

\begin{table}[h]
\begin{centering}
\begin{tabular}{|c|c|c|c|c|c|}
\hline 
$z_{1}$ & $z_{2}$ & $z_{3}$ & $z_{4}$ & $z_{5}$ & $z_{6}$  \tabularnewline
\hline
\hline 
$1$ &$1$ &$1$ &$2$ &$0$ &$1$  \tabularnewline
\hline 
$0$ &$0$ &$0$ &$1$ &$1$ &$1$  \tabularnewline
\hline 
\end{tabular}
\par
\end{centering}
\caption{Projective weights of the coordinates under the two toric $\mathbb{C}^*$ actions for the toric space $T_4/\mathbb{Z}_2$.}
\label{tab:qdp7/z2}
\end{table} 
The SR ideal is 
\begin{equation} \label{srbase}
(z_1 z_2 z_3\,; \, z_4 z_5 z_6)\,.
\end{equation} 
Note, that by orbifolding $T_4$ we lose one row of toric weights. This is exactly analogous to what happens in the treatment of the $\mathbb{P}^1 \times \mathbb{P}^1$ case: The two toric actions of the product space are reduced to the single toric action of $\mathbb{P}^2$.

We may now put the puzzle together. The orbifold of the CY threefold $Q^{(dP_7)^2}$ is given by an equation $P^{(5,2)}(z_i)=0$, of bidegree $(5,2)$\footnote{Throughout the paper, I will often indicate the multi-degree of an equation, as opposed to its corresponding divisor class.}, where the $z_i$'s are the homogeneous coordinates of the quotient toric space $T_4/\mathbb{Z}_2$.

The intersection numbers in the new basis $D_{1}, D_{5}$, where $D_i: z_i=0$, are
\begin{equation} \label{intnumbersb3}
I = 2\,D_{1}^2\,D_{5}-(D_{1}+D_{5})\,D_{5}^2\,.
\end{equation}
The anti-canonical class of this threefold is
\begin{equation}
K^{-1} =  D_{1}+D_{5} > 0\,.
\end{equation}
The anti-canonical class is an effective divisor in the sense that it has a holomorphic representative, i.e. $z_6=0$. However, the anti-canonical bundle in \emph{not} ample. To see this, note that at the locus $z_5=z_6=0$, all sections of the bundle $K^{-n}$ will vanish, for any $n>0$. Therefore, the anti-canonical bundle is not ample, and the threefold is \emph{not} Fano\footnote{In the first version of this paper, I claimed that this threefold is Fano. In light of the results in \cite{Blumenhagen:2009up}, which appeared simultaneously with this paper, I became aware of the contrary. This is also consistent with the no-go theorem of \cite{Cordova:2009fg}.}. We can also compute the properties of the orbifold of the two Del Pezzo surfaces, which is given by
\begin{equation}
P^{(5,2)}(z_i)=0\,, \quad \cap \quad z_5=0\,.
\end{equation}
Its Euler characteristic is $\chi=10$, and its holomorphic Euler characteristic is $\chi_0 = 1$. We can also deduce its $h^{0,2}$ by applying the Serre duality
\begin{equation}
\big(H^{0,2}(D_5, \mathcal{O})\big)^* = H^{0,0}(D_5, K_{B_3} \otimes N_{D_5})\,.
\end{equation}
This amounts to counting monomials of bidegree $(-1,0)$ that are regular and non-vanishing at $z_5=0$, of which there are none in this case. Putting all this information together, we deduce that
\begin{equation}
h^{0,2}=h^{0,1}=0\,, \quad h^{1,1}=8\,.
\end{equation}
This is consistent with a dP$_7$ surface. With some more work, one can algebraically prove that this surface is isomorphic to the two dP$_7$ surfaces $x_6=0$ and $x_7=0$. Hence, the permutation involution has litterally identified the two surfaces.

\section{Constructing CY fourfolds:\\ The quasi-smooth case} \label{sec:quasismooth}
\subsection{Summary of general procedure} \label{subsec: summarygenproc}
In this section, I will summarize the procedure to construct a quasi-smooth\footnote{As explained in \cite{Collinucci:2008zs}, although the fourfolds in this section will not have singularities due to enhancement, they can inherit $\mathbb{Z}_2$-singularities from the base space if O3-planes are present.} \footnote{The term ``quasi-smooth'' will be abused throughout this paper. Because the threefold base is not Fano, the CY fourfold may have singularities that cannot be deformed away. In this context, by ``quasi-smooth'', I will mean a CY fourfold at a generic locus of its complex structure moduli space.} F-theory CY fourfold, given a generic IIB setup with D7-branes and O7/O3-planes without (intended) enhanced gauge groups, by running Sen's weak coupling limit procedure \cite{Sen:1996vd, Sen:1997gv, Sen:1997kw} backwards. See \cite{Denef:2008wq}, for an introduction to this limit. This was explained in detail in \cite{Collinucci:2008zs} for the case of orientifold involutions that act trivially on the homology of the IIB CY threefold, i.e. $h^{1,1}_-(X_3)=0$. However, we will see that the technique can repeated for the case of permutation involutions straightforwardly. The difficult part is finding the right threefold base $B_3$, which was the purpose of the previous chapter. In the next chapter, we will move on to the case of CY fourfolds with enhanced singularities.
\vskip2mm

Let $Y_4$ be a CY fourfold that is elliptically fibered over a threefold $B_3$, with the fibration given by a Weierstrass model
\begin{equation} \label{weierstrass}
y^2 = x^3+x\,f\,z^4+g\,z^6\,,
\end{equation}
where $(x, y, z)$ form a $\mathbb{P}^2_{2,3,1}$, and $f,g$ are sections of appropriate line bundles over $B_3$. Sen's weak coupling limit consists in reparametrizing $f$ and $g$ as follows:
\begin{align} \label{senslimit}
f&= -3\,h^2+\epsilon\, \eta\,,\\
g&=-2\,h^3+\epsilon\,h\,\eta-\frac{\epsilon^2}{12}\,\chi\,, \nonumber
\end{align}
and taking a limit $\epsilon \rightarrow 0$. The resulting configuration turns out to describe an O7-plane located at $h=0$, and a generic D7-brane located at $\eta^2-h\,\chi=0$. The base $B_3$ is then interpreted as the $\mathbb{Z}_2$ quotient of a CY threefold $X_3$, which can be explicitly constructed.

This procedure can be reversed. Given an orientifold model on a CY threefold $X_3$ with an O7-plane at the locus $h=0$, for some polynomial $h$, one can construct the orbifold with the method described in the previous paper, and then deduce the degrees of the polynomials $f$ and $g$ from \eqref{senslimit}, by knowing the degree of $h$. Finally, an ambient toric space can be constructed by augmenting the toric space containing $B_3$ with the coordinates $(x,y,z)$ appropriately. The CY fourfold $Y_4$ is then a complete intersection of the Weierstrass equation and the equations defining $B_3$.

In order to illustrate the general method, this procedure will be carried out explicitly for the CY threefold $Q^{(dP_7)^2}$ in the next section.

\subsection{Lifting a generic configuration on $Q^{(dP_7)^2}$}

In this section, I will explicitly construct an F-theory fourfold describing the most generic D7/O7 configuration on the CY threefold $Q^{(dP_7)^2}$. The non-generic case with enhanced singularities will be dealt with in the next chapter.

We start by choosing our threefold base $B_3$ as the hypersurface given by $P^{(5,2)}=0$, 
in the toric ambient space from table \ref{tab:qdp7/z2}, with intersection numbers \eqref{intnumbersb3}. The locus of the O7 is given by the divisor $h \equiv z_6^2-z_4\,z_5=0$, which has bidegree $(2,2)$, (or alternatively, which is of class $2\,(D_{1}+D_{5})$). By inspecting \eqref{senslimit}, we deduce the degrees of $f$ and $g$
\begin{equation}
{\rm deg}(f) = (4,4) \,, \quad {\rm deg}(g) = (6,6)\,.
\end{equation}
In order for the Weierstrass equation  \eqref{weierstrass} to be well-defined, $(x, y, z)$ must then have degrees
\begin{equation}
{\rm deg}(x) = (2,2)\,, \quad {\rm deg}(y) = (3,3) \,, \quad {\rm deg}(z) = (0,0)\,,
\end{equation}
under the two toric actions of $B_3$. We can now put all this information together, and construct our final ambient space, namely, the one in which our CY fourfold $Y_4$ will live. This is a toric \emph{six}fold $T_6$. Its toric weights are listed in table \ref{tab:ambientsixfold}

\begin{table}[!h] 
\begin{centering}
\begin{tabular}{|c|c|c|c|c|c|c|c|c|}
\hline 
$z_{1}$ & $z_{2}$ & $z_{3}$ & $z_{4}$ & $z_{5}$ & $z_{6}$  &$x$ & $y$ & $z$ \tabularnewline
\hline
\hline 
$1$ &$1$ &$1$ &$2$ &$0$ &$1$  & 2 & 3 & 0 \tabularnewline
\hline 
$0$ &$0$ &$0$ &$1$ &$1$ &$1$  & 2 & 3 & 0\tabularnewline
\hline 
$0$ &$0$ &$0$ & $0$ &$0$ &$0$ & 2 & 3 & 1\tabularnewline
\hline 
\end{tabular}
\par
\end{centering}
\caption{Projective weights of the coordinates under the three toric $\mathbb{C}^*$ actions for the ambient sixfold $T_6$ of the CY fourfold.}
\label{tab:ambientsixfold}
\end{table} 
The SR ideal is given by
\begin{equation} \label{srfourfold}
(z_1\,z_2\,z_3; \, z_4\,z_5\,z_6; \, x\,y\,z)\,.
\end{equation}
The intersection numbers of the CY fourfold are
\begin{align}
I = &2\,D_{1}^2\,(D_{5}\,D_{z}-D_{z}^2)+3\,D_1\,D_z^3-D_1\,D_5^2\,D_z \nonumber\\
&-D_1\,D_5\,D_z^2-D_5\,D_z^3+2\,D_5^2\,D_z^2-D_5^3\,D_z-2\,D_z^4\,.
\end{align}
The Euler characteristic is
\begin{equation}
\chi(Y_4) = 1008\,.
\end{equation}

\subsection{Non-trivial check of tadpole formula}

As a check of the validity of this construction, we can use the formula of Sethi-Vafa-Witten \cite{Sethi:1996es}, which relates the Euler characteristic of the fourfold $\chi(Y_4)$ to the total \emph{curvature}-induced D3 tadpole of a generic D7/O7 configuration as follows:
\begin{equation}
Q_{\rm D3}^{\rm curv.} = \tfrac{1}{12}\,\chi(Y_4) = \tfrac{1}{24}\,\chi({\rm D7})+\tfrac{1}{6}\,\chi({\rm O7})\,,
\end{equation}
where the charges are measured in the `upstairs' convention.
 As explained at great length in \cite{Collinucci:2008pf}, a generic, involution invariant D7-brane will necessarily wrap a singular divisor given by a highly non-transverse equation, 
\begin{equation}
\eta^2-h\,\chi=0\,,
\end{equation}
 where $\eta, h$, and $\chi$ were defined in \ref{subsec: summarygenproc}. This means that we cannot use the standard formulae for Euler characteristics reliably. Let us instead apply the techniques of \cite{Collinucci:2008pf} to compute the curvature-induced D3 charge of such a D7-brane by treating it as a tachyon condensate of two D9/anti-D9 pairs\footnote{As explained in \cite{Uranga:2000xp}, one needs two such pairs in order to cancel the $\mathbb{Z}_2$ K-theory charge.}. 
 
It is easiest to perform this calculation \emph{ in the original CY threefold} $Q^{(dP_7)^2}$, whose toric data are summarized in table \ref{tab:qdp72} and \eqref{intqdp72}. Let us now revert to the old basis $(D_1, D_6, D_7)$, where $D_i: x_i=0$, hoping this will not cause confusion. We can specify the gauge bundle $E$ on the rank two D9 stack as the following Whitney sum:
\begin{alignat}{2}
&{\rm D}9_1\, &\qquad & {\rm D}9_2 \\
&(4-a)\,D_1+(4-b)\,D_6+(4-c)\,D_7 &\quad \oplus \quad& a\,D_1+b\,D_6+c\,D_7\,, \\
\end{alignat}
where $a, b$, and $c$ are arbitrary integers.
The orientifold image anti-D9 stack will then have the following gauge bundle $\sigma^*(E)$
\begin{alignat}{2}
&\overline{{\rm D}9_1}\, &\qquad & \overline{{\rm D}9_2} \\
&(a-4)\,D_1+(c-4)\,D_6+(b-4)\,D_7 &\quad \oplus \quad& -a\,D_1-c\,D_6-b\,D_7\,. 
\end{alignat}
The choice of the three free parameters will determine the gauge bundle that survives on the D7-brane that will result from tachyon condensation. In order to ensure that the tachyon field is a section of a positive bundle, the parameters cannot be arbitrary, but must satisfy the following bounds
\begin{equation}
\tfrac{7}{2} \geq a \geq \tfrac{1}{2}\,, \quad 7 \geq b+c \geq 1\,.
\end{equation}
In order to produce a D7-brane with no DBI flux on it, one must saturate these bounds. The fact that we cannot saturate them with integral parameters may indicate that the D7-brane is wrapping a non-spin cycle, and therefore forcefully carries a DBI flux. Nevertheless, as explained in \cite{Collinucci:2008zs} we can isolate the contribution to $Q_{\rm D3}^{\rm curv.}$ by artificially saturating the bounds with half-integral choices.
Substituting the choice $a=1/2$ and $b+c=1$ into the formula
\begin{equation}
\Big({\rm ch}(E)-{\rm ch}(\sigma^*(E))\Big)\,\sqrt{{\rm Td}(X_3)}\,,
\end{equation}
we get a contribution of $Q_{\rm D3}^{\rm curv.}(D7)=224/3$. The contribution from the O7-plane is easily computed, $Q_{\rm D3}^{\rm curv.}(O7)=56/6$. This gives us a grand total of
\begin{equation}
\fbox{$Q_{\rm D3}^{\rm curv.}(D7)+Q_{\rm D3}^{\rm curv.}(O7) = 84 = \tfrac{1}{12}\,\chi(Y_4)$}
\end{equation}
A perfect match! 

This is a highly non-trivial check that the fourfold $Y_4$ really describes this generic setup. In the next section, we will move on to less generic, but phenomenologically more interesting configurations.

\section{Constructing CY fourfolds:\\ Enhanced singularities} \label{sec:enhanced}
The previous chapter, \ref{sec:quasismooth}, was dedicated to constructing F-theory fourfolds by taking as input data a CY threefold, and an orientifold involution. The D7-brane configuration was assumed to be of the most generic kind. In other words, we saturated the whole D7 tadpole from the O7-plane with a single D7-brane without enhanced gauge group. The procedure resulted in a CY fourfold without any engineered singularity enhancements. That serves as a starting point for less generic, but phenomenologically more interesting configurations with several D7-stacks. 

Moving from the generic case to the case with enhanced singularities consists in taking a (quasi)-smooth F-theory fourfold and moving to regions in its complex structure moduli space where it becomes singular. The goal of this section is to establish a general procedure to accomplish this task.

\subsection{General procedure} \label{subsec:generalcomposition}
One way to engineer the desired singularity enhancements in the fourfold $Y_4$ is to describe it in the  Tate form \cite{Tate, Bershadsky:1996nh}, and to impose that the coefficients of the polynomial have the right orders of vanishing along different loci. There is, however, a more intuitive way to go about creating enhanced singularities corresponding to perturbative brane configurations, by using the Weierstrass form of the fibration. 
\vskip 2mm
Starting with an elliptically fibered CY fourfold $Y_4$, with a Weierstrass model
\begin{equation}
y^2 = x^3+f\,x\,z^4+g\,z^6\,,
\end{equation}
creating a perturbative IIB setup boils down to choosing $f$ and $g$ properly. Sen's Ansatz
\begin{align}
f&= -3\,h^2+\epsilon\, \eta\,,\\
g&=-2\,h^3+\epsilon\,h\,\eta-\frac{\epsilon^2}{12}\,\chi\,,
\end{align}
gives us a system with an O7-plane at $h=0$, and a single, D7-tadpole saturating, D7-brane at 
\begin{equation} \label{whitney1}
\eta^2-h\,\chi=0\,.
\end{equation}
The physical reasons for the peculiar form of the D7-brane were elucidated in \cite{Collinucci:2008pf}. The important feature that characterizes the equation is the fact that the D7-brane intersects the O7-plane at a double curve. This double intersection property was also discussed in \cite{Braun:2008ua}.
\begin{equation}
\eta^2-h\,\chi=0 \quad \cap \quad h=0\qquad  \Rightarrow \eta^2=0\,.
\end{equation}
In order to create different stacks, we need to choose $\eta$ and $\chi$ such that \eqref{whitney1} factorizes into the equations for the different stacks we want to create. Each one of these stacks, taken with their images, must \emph{separately} intersect the O7-plane at a double curve. In other words, if we want a setup with $n$ stacks with respective ranks $r_1,\ldots, r_n$, then we must choose $\eta$ and $\chi$ such that
\begin{equation}
\eta^2 - h\,\chi = (\eta_1^2-h\,\chi_1)^{r_1}\, \ldots \, (\eta_n^2-h\,\chi_n)^{r_n}\,.
\end{equation}

Note the following important composition rule for two rank-one stacks
\begin{equation}
(\eta_1^2-h\,\chi_1)\,(\eta_2^2-h\,\chi_2) = \eta_{\rm tot}^2- h \, \chi_{\rm tot}\,,
\end{equation}
with 
\begin{equation} \label{compositionrule}
\fbox{$\eta_{\rm tot} = \eta_1\,\eta_2\,, \quad {\rm and} \quad \chi_{\rm tot} = \chi_1\,\eta_2^2+\chi_2\,\eta_1^2-h\,\chi_1\,\chi_2\,.$}
\end{equation}
This rule can be used recursively to construct arbitrary configurations. The advantage of using the Weierstrass form of the fibration as opposed to the Tate form, is that this composition method allows one to construct complicated models step by step, retaining the geometric intuition about the D7 stacks, as opposed to trying to guess the right Ansatz in one go. It will be exploited in the next section.

\subsection{Lifting a simple SU(5) model} \label{subsec:simplesu5}
In this section, I will use the method described in \ref{subsec:generalcomposition} to create the F-theory lift for an SU(5) model on the CY threefold $Q^{(dP_7)^2}$, with the permutation involution defined in \ref{subsec:baseqdp72}. The relevant data are repeated here for the reader's convenience. The toric data are in table \ref{tab:qdp72again}. The threefold is a hypersurface given by a polynomial $P^{(5,2,2)}$ of degree $(5,2,2)$. 

\begin{table}[!h] 
\begin{centering}
\begin{tabular}{|c|c|c|c|c|c|c||c|}
\hline 
$x_{1}$ & $x_{2}$ & $x_{3}$ & $x_{4}$ & $x_{5}$ & $x_{6}$ & $x_7$ &$d$\tabularnewline
\hline
\hline 
$1$ &$1$ &$1$ &$1$ &$1$ &$0$ & $0$ & $5$\tabularnewline
\hline 
$0$ &$0$ &$0$ &$0$ &$1$ &$1$ &$0$ &2 \tabularnewline
\hline 
$0$ & $0$ &$0$ &$1$ &$0$& $0$ &$1$ &2 \tabularnewline
\hline 
\end{tabular}
\par
\end{centering}
\caption{Toric weights of the coordinates of the ambient space of the $Q^{(dP_7)^2}$ CY threefold. The last column indicates the degree of a CY hypersurface.}
\label{tab:qdp72again}
\end{table} 
The involution in question is
\begin{equation}
x_4 \leftrightarrow x_5\,, \quad x_6 \leftrightarrow x_7\,,
\end{equation}
which has an O7-plane at 
\begin{equation}
h \equiv x_5\,x_7-x_4\,x_6=0\,.
\end{equation}
The CY fourfold for a generic D7-brane configuration sits in an ambient toric sixfold with data summarized in table \ref{tab:fourfoldagain}

\begin{table}[!h] 
\begin{centering}
\begin{tabular}{|c|c|c|c|c|c|c|c|c|c|}
\hline 
$z_{1}$ & $z_{2}$ & $z_{3}$ & $z_{4}$ & $z_{5}$ & $z_{6}$ &$x$ & $y$ & $z$ \tabularnewline
\hline
\hline 
$1$ &$1$ &$1$ &$2$ &$0$ &$1$  & 2 & 3 & 0 \tabularnewline
\hline 
$0$ &$0$ &$0$ &$1$ &$1$ &$1$  & 2 & 3 & 0\tabularnewline
\hline 
$0$ &$0$ &$0$ & $0$ &$0$ &$0$ & 2 & 3 & 1\tabularnewline
\hline 
\end{tabular}
\par
\end{centering}
\caption{Projective weights of the coordinates under the three toric $\mathbb{C}^*$ actions for the toric space $T_6$.}
\label{tab:fourfoldagain}
\end{table} 
The CY fourfold is a complete intersection of 
\begin{equation} \label{completeint}
P^{(5,2)}=0 \quad \cap \quad y^2=x^3+f\,x\,z^4+g\,z^6\,,
\end{equation}
where $f$ and $g$ have degrees $(4,4,0)$ and $(6,6,0)$, respectively. The coordinates $x_i$ are related to the $z_i$ via the orbifold projection as follows:
\begin{align}
(x_4,\, x_5, \, x_6, \, x_7) \mapsto &(z_4, \, z_5, \,z_6) \nonumber \\
= &\left(2\,x_4\,x_5, \, 2\,x_6\,x_7, \, x_5\,x_7+x_4\,x_6 \right)\,.
\end{align}
\vskip 2mm
We are now ready to choose a scenario. Define a model with the following brane stacks:
\begin{alignat}{2}
{\rm U}(5): \quad & (x_6\,x_7)^5 &=&\, 0 \quad \sim z_5^5=0\,,\\
\text{ unknown gauge group}:\quad & Q^{(8,3,0)}&=& \,0\,, \quad {\rm for \ some} \quad Q^{(8,3,0)}\,.\\
{\rm O}(7)-{\rm plane}: \quad & x_5\,x_7-x_4\,x_6&=& \,0 \quad \sim z_6^2-z_4\,z_5=0\,. \\
\end{alignat}
To build this model, we will proceed stack by stack, and combine the results via the composition rule \eqref{compositionrule}. Let us begin by choosing an $(\eta_1, \chi_1)$ pair to make the U(5) stack. The first thing to notice is that the U(5) stack by itself violates the double-intersection rule, i.e. 
\begin{align*}
z_5^5&=0 \quad \cap \quad h=z_6^2-z_4\,z_5=0 \\ 
\nRightarrow  \eta_1^2&=0\, \qquad \text{for any} \quad \eta_1\,.
\end{align*}
This means that the U(5) stack is not consistent by itself.
To get around this problem, we can take a hint from  the O7-plane equation $z_4\,z_5=z_6^2$. This suggests one simple solution. If we put a U(1) stack on $z_4=0$, then the combined system we are searching for can be written as follows
\begin{align}
\eta_1^2-h\,\chi_1&= z_5^4\,(z_5\,z_4)\,, \qquad \text{which satisfies}\\
z_5^4\,(z_5\,z_4)&=0 \quad \cap \quad h=0 \qquad \Rightarrow\, \eta_1^2=z_5^4\,z_6^2=0\,.
\end{align}
The last equation shows that this system obeys the double-intersection rule, and it tells us how to choose $\eta_1$. The polynomial $\chi_1$ is then easily deduced by making sure that the discriminant has the form $z_5^5\,z_4$. The combined system is then
\begin{equation}
\eta_1 \equiv z_5^2\,z_6\, \quad \chi_1 \equiv z_5^4\,.
\end{equation}
We must now saturate the rest of the D7 tadpole by adding a stack of degree $(6,2,0)$. This means choosing two polynomials $Q$ and $P$ such that
\begin{equation}
\eta_2 = Q^{(3,1,0)}\,, \quad \chi_2 = P^{(4,0,0)}(z_1, z_2, z_3)\,.
\end{equation}
Using the composition rule in \eqref{compositionrule}, we can compute the total polynomials 
\begin{equation}
\eta = z_5^2\,z_6\,Q^{(3,1,0)}\, \quad \chi = z_5^4\,(Q^{(3,1,0)})^2+P^{(4,0,0)}\,z_5^5\,z_4\,.
\end{equation}
We are now ready to write down the full Weierstrass model:
\begin{equation} \label{senansatzu1}
\boxed{
\begin{aligned}
\eta_{\rm tot} &= z_5^2\,z_6\,Q^{(3,1,0)}\, \\ 
\chi_{\rm tot} &= z_5^4\,(Q^{(3,1,0)})^2+P^{(4,0,0)}\,z_5^5\,z_4
\end{aligned}}
\end{equation}
We are not quite finished yet. This choice of $(\eta, \chi)$ is not unique. The discriminant, in Sen's limit, has the following symmetry under changes of the polynomials:
\begin{equation} \label{transfo}
\eta \rightarrow \eta+h\,\psi\,, \quad \chi \rightarrow \chi+2\,\eta\,\psi+h\,\psi^2\,,
\end{equation}
for an arbitrary polynomial $\psi$ of degree $(2,2,0)$. This is then the most general Ansatz that yields the following discriminant:
\begin{equation}
\Delta \sim h^2\,(\eta_{\rm tot}^2-h\,\chi_{\rm tot}) = h^2\,z_5^5\,z_4\,(Q^2-h\,P)\,.
\end{equation}
To summarize, the polynomials in \eqref{senansatzu1} determine the Weierstrass model for the following IIB system:
\begin{equation} \label{scenarioI}
\boxed{
\begin{aligned}
{\rm U}(5)\,: \quad &(x_6\,x_7)^5 =0 & \sim \quad & z_5^5=0 \\
{\rm U}(1)\,: \quad &x_4\,x_5 =0 & \sim \quad & z_4=0 \\
{\rm O}(1)\,: \quad &Q^2-h\,P =0 & \sim \quad &  Q^2-(z_6^2-z_4\,z_5)\,P =0 
\end{aligned}}
\end{equation}
The explanation for the O$(1)$ gauge group can be found in \cite{Collinucci:2008pf}. 
\vskip 3mm
One can now carry out the analysis of singularity enhancements for this Weierstrass model.
For convenience, the classification of fiber singularities relevant here are presented in table \ref{tab:kodaira}
\begin{table}[!h] 
\begin{centering}
\begin{tabular}{|c|c|c|c|}
\hline 
ord(f) & ord(g) & ord($\Delta$) & group\tabularnewline
\hline
$\geq 0$ & $\geq 0$ & $0$ & none\tabularnewline
$0$ & $0$ & $n$ & SU($n$)\tabularnewline
$1$ & $\geq 2$ & $3$ & SU($2$)\tabularnewline
$\geq 2$ & $2$ & $4$ & SU($3$)\tabularnewline
$2$ & $\geq 3$ & $n+6$ & SO($2\,n+8$)\tabularnewline
$\geq 2$ & $3$ & $n+6$ & SO($2\,n+8$)\tabularnewline
$\geq 3$ & $4$ & $8$ & E$_6$\tabularnewline
\hline 
\end{tabular}
\par
\end{centering}
\caption{Classification of enhanced singularities for elliptically fibered CY fourfolds. The first three columns indicate the lowest vanishing order of the polynomial at the locus in question.}
\label{tab:kodaira}
\end{table} 

Applying the classification to our scenario in \eqref{senansatzu1}, one detects the following gauge groups:
\begin{alignat}{5} \label{gaugeggroupsscenarioI}
{\rm locus}\, \qquad & {\rm ord}(f) & \quad & {\rm ord}(g)&\quad &{\rm ord}(\Delta) &\quad  &\text{gauge group} \nonumber \\
z_5=0 \qquad & 0 && 0 &&5 &  &{\rm U}(5) \nonumber \\
z_4=0 \qquad & 0 && 0 &&1 &  &{\rm U}(1) \nonumber \\
z_5=0 \, \cap \, z_4=0 \qquad & 0 && 0 &&6 &  &{\rm U}(6) \\
z_5=0 \, \cap \, z_6=0 \qquad & 2 && 3 &&7 &  &{\rm SO}(10) \nonumber \\
z_5=0 \, \cap \, Q=0 \, \cap \, z_6=0 \qquad & 2 && 3 &&8 &  &{\rm SO}(12) \nonumber\\
\cancel{z_5=0 \, \cap \, z_4=0 \, \cap \, z_6=0} \qquad & 3 && 4 &&8 &  &\cancel{{\rm E}_6} \nonumber 
\end{alignat}
The first five gauge groups are the expected IIB perturbative groups, whereas the last one is inherently non-perturbative, and is much harder to deduce with IIB string theory. 

The first two lines of \eqref{gaugeggroupsscenarioI} correspond to the gauge groups on the gauge stack and the `flavor' stack. The invariant brane at $Q^2-h\,P=0$ only has an O(1) gauge group. 

The third line corresponds to the intersection of the two unitary stacks. The fourth line gives the intersection curve of the U(5) stack with the O7-plane. The fifth line specifies the points where the U(5) stack, the O(1) stack and the O7 meet. Note, that in order to detect this SO(12) gauge group, one must keep the whole expansion in $\epsilon$ of the discriminant $4\,f^3+27\,g^2$. Otherwise, one will incorrectly deduce the lowest vanishing order of $\Delta$.

Finally, the fifth line describes the points where the two unitary stacks and the O7 would meet. However, this locus does not exist, as it lies in the SR ideal \eqref{srbase} . Hence, this model does not have an E$_6$ enhancement, which could have generated a top Yukawa coupling. However, this was to be expected, as the latter is perturbatively forbidden.

\subsection{Lifting a general SU(5) model} \label{subsec:generalsu5}
The solution proposed in \eqref{senansatzu1} solves the problem of finding a pair $(\eta, \chi)$ such that
\begin{equation}
\eta^2-h\,\chi \sim z_5^5\,(\ldots)\,.
\end{equation}
Although it the simplest solution to think of, it is by far not the most general one, and perhaps not the most preferable one for phenomenological reasons. That solution requires the presence of a total of three D7-brane stacks. If one is interested in the stabilization of complex structure moduli, then it is desirable to have as few stacks as possible to saturate the D7-tadpole. The higher the algebraic degree of a D7 stack, the higher its Euler characteristic will be, and therefore, the higher its curvature induced D3-tadpole will be. For instance, in a one-modulus CY, the Euler characteristic of a divisor of degree $N$ grows like $N^3$, whereas the induced D3 charges of a stack of $N$ D7-branes on a degree one divisor goes like $N$.

Therefore, it would be preferable to come up with a solution that has at most two stacks: The U(5)-stack, and one `flavour' stack.
\vskip 2mm
For the time being, we can forget about four of the five branes on $z_5=0$, as these can be easily added in later via the composition rule \eqref{compositionrule}. So the task is to find a polynomial $M^{(8,3,0)}$, such that
\begin{equation}
\eta_1^2-h\,\chi_1 = z_5\,M\,, \quad {\rm and} \quad z_5\,M=0 \, \cap \, h=0\, \Rightarrow z_5\,M=\eta_1^2\,,
\end{equation}
for some $\eta_1$. There are two obvious choices for M that will satisfy this property:
\begin{align}
M_1 &\equiv z_5\,\left(R^{(4,1,0)}\right)^2\,,\quad {\rm satisfies} \quad z_5\,M_1\bigg|_{h=0} = z_5^2\,R^2\,,
\end{align}
and
\begin{align}
M_2 &\equiv z_4\,\left(Q^{(3,1,0)}\right)^2\,,\quad {\rm satisfies} \quad z_5\,M_2\bigg|_{h=0} = z_6^2\,Q^2\,,
\end{align}
where $R$ and $Q$ are arbitrary polynomials of the indicated degrees. The first choice will add a brane to the U(5)-stack, turning it into a U(6)-stack, which we do not want. The second choice is the one that was made in the previous model. However, we can combine both of these choices and complete the square as follows:
\begin{align}
M &= z_5\,\left(R^{(4,1,0)}\right)^2 + z_4\,\left(Q^{(3,1,0)}\right)^2 + 2\,R\,Q\,z_6\,, \quad {\rm so \ that} \\
& z_5\,M\bigg|_{h=0} = (z_5\,R+Q\,z_6)^2\,.
\end{align}
Finally, we can always add a term of the form $h\,S^{(6,1,0)}$ at no extra cost, since it vanishes at the O7-plane. To summarize, we need an Ansatz $(\eta_1, \chi_1)$ such that
\begin{equation} \label{bigansatz}
\eta_1^2-h\,\chi_1 = z_5\,\Big(z_5\,\left(R^{(4,1,0)}\right)^2 + z_4\,\left(Q^{(3,1,0)}\right)^2 + 2\,R\,Q\,z_6-h\,S^{(6,1,0)}\Big)\,.
\end{equation}
This task is accomplished as follows. First, we find $\eta_1$ by intersecting both sides of the equation with the O7-plane. This yields
\begin{equation}
\eta_1 = z_5\,R+Q\,z_6\,.
\end{equation}
Then, we can easily deduce $\chi_1$ by substituting this expression for $\eta_1$ in \eqref{bigansatz}. This yields 
\begin{equation}
\chi_1 = Q^2+z_5\,S\,.
\end{equation}
We may now combine this with the Ansatz, $(\eta_2, \chi_2) = (z_5^2, 0)$, for the other four branes on $z_5=0$ with the composition rule \eqref{compositionrule}. Finally, we obtain the following complete form of the Weierstrass model:
\begin{equation} \label{senansatzgeneral}
\boxed{
\begin{aligned}
\eta_{\rm tot} &= z_5^3\,R^{(4,1,0)}+z_5^2\,z_6\,Q^{(3,1,0)}\, \\ 
\chi_{\rm tot} &=z_5^4\,\left(Q^{(3,1,0)}\right)^2+z_5^5\,S^{(6,1,0)}\,.
\end{aligned}}
\end{equation}
We can be even more general if we take into account the transformation rule \eqref{transfo} that leaves the discriminant invariant in Sen's limit. However, this will not be very helpful for the purpose of detecting singularities. The discriminant is
\begin{equation} \label{generaldiscriminant}
\fbox{$\Delta \sim h^2\,z_5^5\,(Q^2\,z_4+R^2\,z_5+2\,R\,Q\,z_6-h\,S)$}
\end{equation}
Defining 
\begin{equation}
K \equiv Q^2\,z_4+R^2\,z_5+2\,R\,Q\,z_6-h\,S\,,
\end{equation}
the IIB setup described by this model is summarized as follows:
\begin{equation} \label{generalscenario}
\boxed{
\begin{aligned}
{\rm U}(5)\,: \quad & z_5^5=0 \\
{\rm O(1)}\,: \quad &K=0 \,.
\end{aligned}}
\end{equation}
where the second stack invariant under the involution. Actually more work is needed to determine the gauge symmetry that survives on the second stack. This is best done by using the K-theoretic techniques laid out in \cite{Collinucci:2008pf}, however, we will not need to determine it here.
\vskip 3mm
We can now study the enhanced singularities of this Weierstrass model:
\begin{alignat}{5} \label{gaugeggroupsscenarioI}
{\rm locus}\, \qquad & {\rm ord}(f) & \quad & {\rm ord}(g)&\quad &{\rm ord}(\Delta) &\quad  &\text{gauge group} \nonumber \\
z_5=0 \qquad & 0 && 0 &&5 &  &{\rm U}(5) \nonumber \\
K=0 \qquad & 0 && 0 &&1 &  &{\rm none} \nonumber \\
z_5=0 \, \cap \, K=0 \qquad & 0 && 0 &&6 &  &{\rm U}(6) \\
z_5=0 \, \cap \, z_6=0 \qquad & 2 && 3 &&7 &  &{\rm SO}(10) \nonumber \\
z_5=0 \, \cap \, Q=0 \, \cap \, z_6=0 \qquad & 2 && 3 &&8 &  &{\rm SO}(12) \nonumber\\
\cancel{z_5=0 \, \cap \, z_4=0 \, \cap \, z_6=0} \qquad & 3 && 4 &&8 &  &\cancel{{\rm E}_6} \nonumber \\
\end{alignat}
The last two lines require some explanation. A priori, the matter curve $z_5=K=0$ intersects the O7-plane at two different sets of points:
\begin{equation}
z_5=0\, \cap \, K=0 \, \cap z_6=0 \quad \Rightarrow \quad z_4\,Q^2=0\,.
\end{equation}
However, the locus where $(z_4, z_5, z_6)$ vanish does not exists, since it lies in the SR ideal \eqref{srbase}. Hence, there is no E$_6$ enhancement, as expected for a perturbative setup.
Note, that just as in the previous model, in order to accurately detect the gauge groups, one must keep the whole expansion in $\epsilon$ of the discriminant $4\,f^3+27\,g^2$.
\vskip 3mm
Note, that the model in \eqref{senansatzu1} is recovered by setting
\begin{equation}
R \rightarrow 0\,, \quad S \rightarrow z_4\,P\,.
\end{equation}

It is useful to relate this Weierstrass model to the Tate form \cite{Tate, Bershadsky:1996nh}. Following \cite{Donagi:2009ra}, 
we can rewrite the equation for the elliptic fibration in the following form:
\begin{equation}
y^2+a_1\,x\,y\,z+a_3\,y\,z^3= x^3+a_2\,x^2\,z^2+a_4\,x\,z^4+a_6\,z^6\,.
\end{equation}
where $a_i$ has degree $(i, i, 0)$. Note, that in the literature one will usually see this equation without the coordinate `$z$'. I include it here, as it is necessary in order to define a compact fibration. Using the transformation from the Weierstrass to the Tate form in \cite{Donagi:2009ra}, we can deduce that the $a_i$'s in this model have the following form:
\begin{equation}
a_1 \sim z_6\,, \quad a_2 \sim z_5\,z_4\,, \quad a_3 \sim z_5^2\,Q\,, \quad a_4 \sim z_5^3\,R\,, \quad a_6 \sim z_5^5\,S\,.
\end{equation}
From this form, one can corroborate all the gauge groups found above from the Weierstrass form. The advantage of the Weierstrass form, however, is that it allows us to construct specific models where there are several stacks in a more intuitive way, as illustrated in section \ref{subsec:simplesu5}.

\section{Conclusions}
In this paper two steps were taken to advance the program of constructing F-theory lifts of global, perturbative IIB O7/O3 models. 

First, a method was devised to construct threefolds as orbifolds of CY threefolds with respect to permutations involutions that act non-trivially on the homology of the threefolds, i.e. $h^{1,1}_- \neq 0$. The case with $h^{1,1}_- = 0$ was treated in \cite{Collinucci:2008zs}. These threefolds serve as base manifolds for elliptically fibered, compact CY fourfolds. The method was applied to a specific example.

The validity of the construction of the specific fourfold was successfully tested in a highly non-trivial way by computing the curvature induced D3 charge of a generic setup both with the Sethi-Vafa-Witten formula, and with the K-theoretic methods of \cite{Collinucci:2008pf}.

Once the CY fourfold was known in its generic form, it was taken to loci in the complex structure moduli space with enhanced singularities. I showed a general procedure to lift SU(5) setups in an intuitive way by using the Weierstrass form of the fibration. The procedure allows one to construct D7-brane setups stack by stack. The construction was tested by identifying the expected enhanced gauge groups at intersections. As expected, only perturbative gauge groups were found, and the E$_6$ locus turned out to be torically excluded from the space.
Finally, a more complicated model was lifted in \ref{app}.

The methods in this paper should open up the possibility to test all kinds of properties of IIB string theory setups. For instance, the problem of zero-mode counting of Euclidean D3-instantons can now be formulated into Witten's criterion \cite{Witten:1996bn}.

It would be interesting to be able to address the problem of explicitly constructing four-form fluxes in these models. This is left for future work.

\bigskip

\begin{center}
{\bf Acknowledgements}\\
\end{center}
I am grateful to Ralph Blumenhagen, Christoph Mayrhofer, and Dennis Westra for useful discussions. This work is supported by the Austrian Research Funds FWF under grant number P19051-N16.

\appendix 
\section{An SU(5) model on $Q^{(dP_9)^2}$} \label{app}
The series of CY threefolds constructed in \cite{Blumenhagen:2008zz} via Del Pezzo transitions from the quintic CY are particularly nice to work with because they can be defined sequentially. One starts out with the quintic, and proceeds to blow-up toric points in the ambient space. With each subsequent blow-up, a dP surface is introduced, and the degree of all surfaces is raised, i.e. dP$_i \rightarrow$ dP$_{i+1}$. The surfaces intersect pairwise.

In this appendix, I will outline how to apply the techniques shown in this paper to construct the F-theory lift of a specific (three-generation) model from \cite{Blumenhagen:2008zz} on $Q^{(dP_9)^4}$, which has four dP$_9$'s. The toric weights of the coordinates of the ambient space for $Q^{(dP_9)^4}$ are shown in \ref{tab:qdp94}.

\begin{table}[!h] 
\begin{centering}
\begin{tabular}{|c|c|c|c|c|c|c|c|c||c|}
\hline 
$x_{1}$ & $x_{2}$ & $x_{3}$ & $x_{4}$ & $x_{5}$ & $x_{6}$ & $x_7$ & $x_{8}$ & $x_9$ &$d$\tabularnewline
\hline
\hline 
$1$ &$1$ &$1$ &$1$ &$1$ &$0$ & $0$ &$0$ & $0$ & $5$\tabularnewline
\hline 
$0$ &$0$ &$0$ &$0$ &$1$ &$1$ &$0$ &$0$ & $0$ &2 \tabularnewline
\hline 
$0$ & $0$ &$0$ &$1$ &$0$& $0$ &$1$ &$0$ & $0$ &2 \tabularnewline
\hline 
$0$ & $0$ &$1$ &$0$ &$0$& $0$ &$0$ &$1$ & $0$ &2 \tabularnewline
\hline 
$0$ & $1$ &$0$ &$0$ &$0$& $0$ &$0$ &$0$ & $1$ &2 \tabularnewline
\hline
\end{tabular}
\par
\end{centering}
\caption{Toric weights for the coordinates of the ambient fourfold of the $Q^{(dP_9)^4}$ CY threefold. The last column indicates the degree of a CY hypersurface. Note, however, that the Mori cone for this space is not simplicial.}
\label{tab:qdp94}
\end{table} 
In the basis $\{D_1, D_i\}, i=6,\,\ldots,\, 9$ The intersection numbers are 
\begin{equation} \label{intqdp94}
I = D_1\,\left(\sum_{i\neq j} D_i\,D_j \right)-\sum_{i}\, D_i^2\,\left(\sum_{j \neq i} D_j \right)-D_1^3\,.
\end{equation}

The orientifold involution is the same as in \ref{subsec:baseqdp72}, i.e. $x_4 \leftrightarrow x_5\,;\, x_6\leftrightarrow x_7$. Hence, the new dP's are not involved, and we can easily construct the generic fourfold $Y_4$ as before. The coordinates with respective toric weights for the ambient sixfold are in \ref{tab:fourfoldqdp94}.

\begin{table}[!h] 
\begin{centering}
\begin{tabular}{|c|c|c|c|c|c|c|c|c|c|c|c|}
\hline 
$z_{1}$ & $z_{2}$ & $z_{3}$ & $z_{4}$ & $z_{5}$ & $z_{6}$ & $z_{7}$ & $z_{8}$ &$x$ & $y$ & $z$ \tabularnewline
\hline
\hline 
$1$ &$1$ &$1$ &$2$ &$0$ &$1$ &$0$ & $0$ & $2$ & $3$ & 0\tabularnewline
\hline 
$0$ &$0$ &$0$ &$1$ &$1$ &$1$ &$0$ & $0$ &2 & $3$ & 0 \tabularnewline
\hline 
$0$ & $0$ &$1$ &$0$ &$0$& $0$ &$1$ & $0$ &2 & $3$ & 0 \tabularnewline
\hline 
$0$ & $1$ &$0$ &$0$ &$0$& $0$ &$0$ & $1$ &2 & $3$ & 0 \tabularnewline
\hline 
$0$ & $0$ &$0$ &$0$ &$0$& $0$ &$0$ & $0$ &2 & $3$ & 1 \tabularnewline
\hline 
\end{tabular}
\par
\end{centering}

\caption{Toric weights for the coordinates of the ambient sixfold of the CY fourfold $Y_4$.}
\label{tab:fourfoldqdp94}
\end{table} 
One portion of the intersection numbers of $Y_4$ is simply $I\times D_{z}$, where $I$ is \eqref{intqdp94}, and the rest can be deduced by adding the element $(x, y, z)$ to the SR ideal.

The specific three-generation model in \cite{Blumenhagen:2008zz} has the following stacks:
\begin{equation} \label{scenariou3}
\boxed{
\begin{aligned}
A: \quad {\rm U}(5)\,: \quad & z_5^5=0 \\
B:\quad {\rm U}(3)\,: \quad & z_4^3=0 \\
C: \quad {\rm U}(1)\,: \quad & z_1^2 =0
\end{aligned}}
\end{equation}
The last stack is actually a double stack on $x_1=0$ in the original $X_3$, so one might expect the gauge group to be USp$(2)$. However, since the cycle is not spin, it has a half-integer, diagonal flux of the form diag$(F, -F)$ that breaks the USp$(2)$ to U(1).
\vskip 2mm
The Weierstrass model can be readily inferred by means of the general methods of section \ref{sec:enhanced}. More, specifically, from \eqref{generaldiscriminant} we see that by setting
\begin{equation}
Q \equiv z_4\,z_1\,, \quad R \equiv 0\,, \quad S \equiv 0\,,
\end{equation}
we get the general Ansatz, (modulo the transformation \eqref{transfo}):
\begin{equation} \label{senansatzu3}
\boxed{
\begin{aligned}
\eta_{\rm tot} &= z_5^2\,z_6\,z_4\,z_1\, \\ 
\chi_{\rm tot} &= z_5^4\,z_4^2\,z_1^2
\end{aligned}}
\end{equation}
which yields the discriminant
\begin{equation}
\Delta \sim h^2\,z_5^5\,z_4^3\,z_1^2\,.
\end{equation}
The new interesting enhanced locus is:
\begin{align}
A &\cap C \cap {\rm O}7: {\rm SO}(14)\,,
\end{align}
and the locus $A \cap B \cap$ O7 does not exist.
\bibliographystyle{utphys}
\bibliography{liftagain}

\end{document}